\documentclass{PoS}

\usepackage{graphicx}
\usepackage[footnotesize,bf]{caption}
\usepackage{arydshln}

\usepackage{xcolor}
\colorlet{alert}{red!60!black}
\colorlet{example}{green!60!black}
\colorlet{structure}{blue!60!black}

 \newcommand{\note}[1]{}                    % Suppress notes in draft

\newcommand{\ind}[1]{\rm\scriptscriptstyle #1}

\def\lsim{\mathrel{\rlap{\lower4pt\hbox{\hskip1pt$\sim$}}
    \raise1pt\hbox{$<$}}}                % less than or approx. symbol
\def\gsim{\mathrel{\rlap{\lower4pt\hbox{\hskip1pt$\sim$}}
    \raise1pt\hbox{$>$}}}                % greater than or approx. symbol

\title{
\vspace*{-0.75cm}
\begin{minipage}{\textwidth}
\begin{flushright}
\texttt{\footnotesize
PoS(LATTICE2019)220   \\
ADP-20-4/T1114        \\
DESY 20-009           \\
Liverpool LTH 1224    \\
}
\end{flushright}
\end{minipage}\\[15pt]
\vspace*{+0.75cm}
        Determining the glue component of the nucleon}

\ShortTitle{Determining the glue component \ldots}

\author{\speaker{R.~Horsley}$^{\,a}$,
        T.~Howson$^b$, 
        W.~Kamleh$^b$,
        Y.~Nakamura$^c$,
        H.~Perlt$^d$,
        P.~E.~L. Rakow$^e$,
        G.~Schierholz$^f$,
        H.~St\"uben$^g$,
        R.~D.~Young$^b$
        and J.~M. Zanotti$^b$ \\
        \llap{$^a$} School of Physics and Astronomy,
                    University of Edinburgh,
                    Edinburgh  EH9 3FD, UK \\
        \llap{$^b$} CSSM, Department of Physics,
                    University of Adelaide, Adelaide SA 5005, Australia \\
        \llap{$^c$} RIKEN Advanced Institute for Computational Science,
                    Kobe, Hyogo 650-0047, Japan \\
        \llap{$^d$} Institut f\"ur Theoretische Physik,
                    Universit\"at Leipzig, 04109 Leipzig, Germany \\
        \llap{$^e$} Theoretical Physics Division,
                    Department of Mathematical Sciences,
                    University of Liverpool, \\
        \llap{\phantom{$^e$}} 
                    Liverpool L69 3BX, UK \\
        \llap{$^f$} Deutsches Elektronen-Synchrotron DESY,
                    22603 Hamburg, Germany \\
        \llap{$^g$} Universit\"at Hamburg, Regionales Rechenzentrum,
                    20146 Hamburg, Germany \\
        E-mail: \email{rhorsley@ph.ed.ac.uk} }

\author{QCDSF-UKQCD-CSSM Collaborations}

\abstract{Computing the gluon component of momentum in
          the nucleon is a difficult and computationally
          expensive problem, as the matrix element involves a
          quark-line-disconnected gluon operator which suffers
          from ultra-violet fluctuations. But also necessary for
          a successful determination is the non-perturbative
          renormalisation of this operator. As a first step
          we investigate here this renormalisation in the $RI-MOM$
          scheme. Using quenched QCD as an example, a statistical signal 
          is obtained in a direct calculation using an 
          adaption of the Feynman-Hellmann technique.}

\FullConference{37th International Symposium on Lattice Field Theory 
                - Lattice2019  \\
		16-22 June 2019\\
		Wuhan, China}

\begin{document}

%----------------------------------------------------------------------------

\section{Introduction}

%----------------------------------------------------------------------------

How the nucleon's momentum is distributed among its constituents is a
question that has been discussed for many years. Indeed the fact that 
the measurement of the fraction of the nucleon momentum carried by
quarks did not sum up to one gave early indications for the existence of
the gluon and QCD. If $\langle x \rangle_f$ is the fraction of nucleon 
momentum carried by parton $f$ (quark, $q$, or gluon, $g$) then we have
\begin{eqnarray}
   \sum_q \langle x \rangle_q + \langle x \rangle_g = 1 \,,
\label{sum_rule}
\end{eqnarray}
where experimentally $\langle x \rangle_g \sim \textstyle{1 \over 2}$.
This talk will describe our progress in a determination of the
renormalisation of $\langle x \rangle_g$ using lattice gauge theory 
techniques. Previous work includes
  \cite{Gockeler:1996zg,Meyer:2007tm,Horsley:2012pz,
        Alexandrou:2016ekb,Yang:2018nqn,Shanahan:2018pib}.
The aim here will be to compare with the previous QCDSF-UKQCD result 
\cite{Horsley:2012pz} but now using the Feynman--Hellmann (FH) theorem to 
also determine the $Z_g$ renormalisation constant using the $RI-MOM$ 
renormalisation procedure, \cite{Chambers:2014pea}, rather than
imposing the sum rule, eq.~(\ref{sum_rule}). 

The relevant operators that we consider here ($\langle x \rangle \equiv v_n$
with $n=2$, for both quark/gluon) are 
\begin{eqnarray}
   {\langle N(\vec{p}) | \widehat{\cal O}^{(b)}_f | N(\vec{p}) \rangle
     \over \langle N(\vec{p}) | N(\vec{p}) \rangle}
        = - {4 \over 3} E \left( 1 - {m^2 \over 4E^2} \right)
                                                   \langle x \rangle_f \,,
   \quad \mbox{where} \quad
   {\cal O}^{(b)}_f = O_{44}^{(f)} - \textstyle{1 \over 3} O_{ii}^{(f)} \,,
\end{eqnarray}
and with%
\footnote{We have slightly changed our convention for $O_{\mu\nu}^{(g)}$
compared to \cite{Horsley:2012pz}.}
\begin{eqnarray}
   O_{\mu\nu}^{(g)} & = F^a_{\mu\alpha} F^a_{\nu\alpha} \,,
                 & \qquad O^{(b)}_g = \textstyle{2 \over 3}
                                     (E_i^{a\,2} - B_i^{a\,2})\,,
                                                           \nonumber \\
   O_{\mu\nu}^{(q)} & = \bar{q}\gamma_\mu
                       \stackrel{\leftrightarrow}{D}_\nu q \,,
                 & \qquad O^{(b)}_q = \bar{q}\gamma_4
                                      \stackrel{\leftrightarrow}{D}_4 q
                                     - \textstyle{1 \over 3}
                                      \bar{q}\gamma_i
                                        \stackrel{\leftrightarrow}{D}_i q \,,
\end{eqnarray}
(using the Euclidean metric) with the notation 
${\cal O}(\tau) = \sum_{\vec{x}}O(\tau,\vec{x})$ with
$\stackrel{\leftrightarrow}{D} 
= (\stackrel{\leftarrow}{D} - \stackrel{\rightarrow}{D})/2$.
This representation for the gluon, $O^{(b)}_g$, allows for 
$\vec{p} = \vec{0}$ in the above, while for the other representation
$O^{(a)}_g \sim \vec{E}\times\vec{B}$, but now $\vec{p} = \vec{0}$ 
is not possible. $\langle x \rangle_g$ is related (and equivalent) 
to the decomposition of the nucleon mass via the energy--momentum tensor, 
\cite{Ji:1994av}. 
As $O^{(b)}_f = \textstyle{4 \over 3}\bar{T}_{44}^{(f)}$ where
$\bar{T}_{\mu\nu}$ is the traceless energy-momentum tensor, then
for example the gluon contribution to the nucleon mass, 
$m$, is $\sim \textstyle{3 \over 4} m\langle x \rangle_g$. As is well known
this can be generalised and used (for higher $n$) in the OPE, for example
for DIS.

%----------------------------------------------------------------------------

\section{Lattice $\langle x \rangle_g$}

%----------------------------------------------------------------------------

Rather than forming ratios of $3$-point to $2$-point correlation
functions which are very noisy, \cite{Gockeler:1996zg}, we choose instead
to add the operator of interest to the action, \cite{Horsley:2012pz}
\begin{eqnarray}
   S \to S(\lambda) = S + \lambda \sum_\tau {\cal O}(\tau) \,,
\end{eqnarray}
and perform subsidiary runs at different $\lambda$s. $E(\lambda)$ is 
then determined and the Feynman--Hellmann (FH) theorem is then 
used to find the matrix element of interest
\begin{eqnarray}
   \left. {\partial E(\lambda) \over \partial \lambda} \right|_{\lambda = 0}   
      =  {\left\langle N \left| : \widehat{\cal O} :
                                \right| N \right\rangle 
           \over \langle N | N \rangle } \,,
\end{eqnarray}
(where $:\ldots:$ means that the vacuum term has been subtracted).
For quark operators this method includes both quark-line -connected 
and -disconnected terms. We shall illustrate here the purely 
quark-line-disconnected 
$\langle x \rangle_g^{\ind dis} \equiv \langle x \rangle_g$ for quenched QCD.

Using the Wilson gluonic action 
as $\mbox{Re}\,\mbox{tr}_{\ind C} [1 - U_{\mu\nu}^{\ind plaq}(x) ]
= \textstyle{1 \over 4} a^4 g^2 F_{\mu\nu}^a(x)^2 + \ldots$
motivates the simplest definition of electric and magnetic fields on 
each time slice as
\begin{eqnarray}
   \textstyle{1 \over 2}{\cal E}^{a\,2}(\tau)
      &=& \textstyle{1 \over 3}\beta \sum_{\vec{x}\,i}\,
                               \mbox{Re}\,\mbox{tr}_c
                    \left[1 - U_{i4}^{\ind plaq}(\vec{x},\tau) \right] \,,
                                                         \nonumber \\
          \textstyle{1 \over 2}{\cal B}^{a\,2}(\tau)
      &=& \textstyle{1 \over 3}\beta \sum_{\vec{x}\,i<j}\,
                               \mbox{Re}\,\mbox{tr}_c
                    \left[1 - U_{ij}^{\ind plaq}(\vec{x},\tau) \right] \,,
\end{eqnarray}
($\beta = 6/g^2$). The modified action in this case is
\begin{eqnarray}
   S(\lambda)
      = \sum_\tau \left( \textstyle{1 \over 2}
             [{\cal E}^{a\,2}(\tau) + {\cal B}^{a\,2}(\tau)]
                        + \lambda \textstyle{3 \over 4} {\cal O}^{(b)}(\tau) 
                 \right) \,,
\label{S_lambda}
\end{eqnarray}
with 
${\cal O}^{(b)}(\tau) = \textstyle{2 \over 3} [{\cal E}^{a\,2}(\tau) 
                                              - {\cal B}^{a\,2}(\tau)]$.
This can be implemented by generating anisotropic lattices.
In \cite{Horsley:2012pz} we have described the determination of 
$\langle x \rangle_g^{\ind lat}$ using this method.

%----------------------------------------------------------------------------

\section{Renormalisation}

%----------------------------------------------------------------------------

We now discuss some aspects of our renormalisation procedure, the main
goal of this talk. We shall only consider here renormalisation for the 
quenched case, \cite{Gockeler:1996zg,Meyer:2007tm} 
-- in the conclusion and outlook section we shall comment on 
the case when dynamical quarks are included.

%----------------------------------------------------------------------------

\subsection{General considerations}

%----------------------------------------------------------------------------

We expect the renormalisation pattern to be for the gluon and
(two) valence quarks
\begin{eqnarray}
   \left( \begin{array}{c}
             \langle x \rangle_g      \\
             \langle x \rangle^{\ind con}_u   \\
             \langle x \rangle^{\ind con}_d   \\
          \end{array}
    \right)^{\ind R}
    = \left( \begin{array}{ccccc} 
                Z_{gg} & Z_{gq} & Z_{gq} \\
                  0   & Z_{qq} & 0     \\
                  0   & 0     & Z_{qq} \\
              \end{array}
      \right) \,
      \left( \begin{array}{c}
             \langle x \rangle_g      \\
             \langle x \rangle^{\ind con}_u   \\
             \langle x \rangle^{\ind con}_d   \\
          \end{array}
    \right)^{\ind lat} \,.
\end{eqnarray}
where $^{\ind con}$, (connected) or valence here means only for 
quark-line connected terms in the correlation function.
In the quenched limit, we have no disconnected quark-line
terms, so we shall drop this index here.
For the bottom two rows of the renormalisation matrix, the zeroes 
are justified because if you don't put in a valence $\langle x \rangle_q$
`by hand' then it remains zero. 

Due to the momentum sum rule, we must have 
\begin{eqnarray}
   \left( \langle x \rangle_g + \langle x \rangle_u
                              + \langle x \rangle_d \right)^{\ind R}
    = Z_g \langle x \rangle_g^{\ind lat}
                   + Z_q \left( \langle x \rangle_u
                               +\langle x \rangle_d \right)^{\ind lat}
    = 1 \,,
\label{mom_sum}
\end{eqnarray}
where $Z_g$, $Z_q$ just depend on the coupling (and so in the quenched 
limit does $Z_{gg}$). Hence we have
\begin{eqnarray}
   Z_g = Z_{gg}\,, \qquad
   Z_q = Z_{gq}^{\ind \overline{MS}} + Z_{qq}^{\ind \overline{MS}} \,.
\end{eqnarray}
We now discuss our procedure for estimating $Z_g$ from $RI-MOM$ and FH.
The standard procedure is used here for $RI-MOM$. We first define the $2$- 
and $1$-particle-irreducible (or $1PI$) correlation functions, 
$D_\lambda$, $\Gamma^{(b)}(p)$ respectively, as 
$\langle A(p) A(-p) \rangle_\lambda = D_\lambda(p)$ and
\begin{eqnarray}
   \langle A(p) O^{(b)} A(-p) \rangle_0 
      = - \left. {4 \over 3} {\partial \over \partial \lambda} 
                   D_\lambda(p) \right|_{\lambda=0}
      = D_0(p) \Gamma^{(b)}(p) D_0(p) \,.
\label{D+Gam_def}
\end{eqnarray}
We expect their structures to be of the form
\begin{eqnarray}
   D_0(p) = D_0^{\ind Born}(p) \, \Delta_0(p^2)\,, \qquad
   \Gamma^{(b)}(p) = \Gamma^{(b)\,{\ind Born}}(p) \,\Lambda^{(b)}(p^2) \,,
\label{int_D+Gam}
\end{eqnarray}
where $D_0^{\ind Born}(p)$, $\Gamma^{(b)\,{\ind Born}}(p)$ are the tree level
or Born terms. The renormalisation constants are specified by
\begin{eqnarray}
   A^{\ind R} = Z_3^{1/2} A \quad\mbox{and}\quad O^{(b)\,\ind R} = Z_g O^{(b)}
   \qquad\Rightarrow\quad 
      D_0^{\ind R} = Z_3 D_0 \,, \quad
      \Gamma^{(b)\,{\ind R}} = Z_gZ_3^{-1} \Gamma^{(b)} \,.
\end{eqnarray}
To define $Z_3$, $Z_g$ we take the renormalisation conditions as
\begin{eqnarray}
   \left. 
      \begin{array}{cc}
         \left. D_0^{\ind R}(p) \right|_{p^2 = \mu^2}
            = \left. D_0^{\ind Born}(p)\right|_{p^2 = \mu^2} 
                                         \\[0.5em]
         \left. \Gamma^{(b)\,\ind{R}}(p) \right|_{p^2 = \mu^2}
            =  \left. \Gamma^{(b)\,\ind{Born}}(p) 
                      \right|_{p^2 = \mu^2} \\
      \end{array}
      \right\}
      \quad\Rightarrow\quad
         Z_3 = {1 \over \Delta_0} \,,
              \quad Z_g = {1 \over \Lambda^{(b)} \Delta_0} \,.
\label{ren_cond}
\end{eqnarray}
So effectively we have to determine $\Delta_0$, $\Lambda^{(b)}$.
This thus first necessitates a determination of the Born correlation functions.

%----------------------------------------------------------------------------

\subsection{The Born correlation functions}
\label{born}

%----------------------------------------------------------------------------

After some algebra, we find that the Born propagator for arbitrary $\lambda$ 
and general gauge fixing parameter, $\xi$, is given by
\begin{eqnarray}
   D^{\ind Born}_\lambda(p)^{ab}_{\mu\nu}
      = \left( {a_{\mu\nu} \over p^2+\lambda(p_4^2-\vec{p}^2)}
               + { b_{\mu\nu} \over (1+\lambda)p^2 }
               + \xi {c_{\mu\nu} \over p^2} \right) \delta^{ab} \,,
\label{D_lam}
\end{eqnarray}
where
\begin{eqnarray}
   a_{\mu\nu} = \delta_{\mu\nu} - {p_\mu p_\nu \over p^2}
                            - {b_\mu b_\nu \over b^2} \,, \qquad
   b_{\mu\nu} = {b_\mu b_\nu \over b^2} \,, \qquad
   c_{\mu\nu} = {p_\mu p_\nu \over p^2} \,,
\label{abs_def}
\end{eqnarray}
and $b = (\vec{p}p_4,-\vec{p}^{\,2})$. Note that $b_\mu$ thus satisfies
$b\cdot p = 0$ and $b^2 = p^2\vec{p}^{\,2}$. Furthermore $a$, $b$ and $c$ 
are orthogonal projectors, which simplifies calculations considerably.
Using $D^{{\ind Born} \,-1}_\lambda$, which is well defined and can be 
immediately found from eq.~(\ref{D_lam}) gives upon generalising 
the definition in eq.~(\ref{D+Gam_def}) to arbitrary $\lambda$,
\begin{eqnarray}
   \Gamma^{(b)\,\ind Born}(p)^{ab}_{\mu\nu}
    &=& {4 \over 3}\left[ a_{\mu\nu} (p_4^2 - \vec{p}^2) 
                          + p^2 b_{\mu\nu} \right]\delta^{ab} \,,
\end{eqnarray}
which is independent of $\lambda$ and also independent of $\xi$.

%----------------------------------------------------------------------------

\subsection{Renormalisation conditions}

%----------------------------------------------------------------------------

We are now in a position to compute $\Delta_0$, $\Lambda^{(b)}$ from
eq.~(\ref{int_D+Gam}) and hence $Z_g$ from eq.~(\ref{ren_cond}). Using 
the results of section~\ref{born} and eq.~(\ref{int_D+Gam}) 
we have the equations
\begin{eqnarray}
   D_0(p) = D_0^{\ind{Born}}(p)\Delta_0(p^2)\,, \quad
   - {4 \over 3} \left. {\partial \over \partial \lambda} 
                      D_\lambda(p) \right|_{\lambda=0}
      = {1 \over (p^2)^2}
          \Lambda^{(b)}(p^2) \, \Delta_0(p^2)^2 \, \Gamma^{(b)\, \ind{Born}}(p) \,.
\end{eqnarray}
There are now many possibilities. We can simply take the trace
of these equations. This gives
\begin{eqnarray}
   Z_g = \left. {1 \over 3} \left( 1 - 4{p_4^2 \over p^2}\right)
         { \mbox{tr}\, D_0(p) \over 
               \left. {\partial \over \partial \lambda}
                     \mbox{tr}\, D_\lambda(p) \right|_{\lambda=0} } 
               \right|_{p^2 = \mu^2} \,,
\label{Z_g_I}
\end{eqnarray}
(where $\mbox{tr}\, X \equiv X_{\mu\mu}^{aa}$). 
%{\color{red}A potential problem is that around $p^2 \sim 4p_4^2$, 
%the expression is singular, we have a form of $0/0$. Close to this it 
%may be noisy.}
Another possibility might be to first multiply by $\Gamma^{(b)\,\ind Born}$ 
before taking the trace. This gives
\begin{eqnarray}
   Z_g = \left. 
         {4 \over 3} {1 \over 3} 
          \left( 1 + 2\left(1-2{p_4^2\over p^2}\right)^2\right)
         { p^2 \mbox{tr}\, D_0(p) \over 
               \left. {\partial \over \partial \lambda}\mbox{tr}\,
                     D_\lambda(p) \Gamma^{(b)\,\ind Born} \right|_{\lambda=0} } 
        \right|_{p^2 = \mu^2} \,.
\label{Z_g_II}
\end{eqnarray}
%{\color{red}which no longer has a singularity.}

%----------------------------------------------------------------------------

\subsection{Preliminary results}

%----------------------------------------------------------------------------

Practically to reduce lattice artifacts, if the gluon propagator is 
defined in the natural way, with distances measured from the mid-point 
of the link, i.e. 
$A_\mu(x+\hat{\mu}/2) = (1/2ig) ( (U_\mu(x) - U_\mu(x)^\dagger) 
               - \mbox{tr}_{\ind C}(U_\mu(x) - U_\mu(x)^\dagger) )$ and
$A^a_\mu(p) = \sum_x e^{ip\cdot (x+\hat{\mu}/2)} A_\mu^a(x+\hat{\mu}/2)$,
which is important if $\mu \not= \nu$, then we get the tree-level results
by the substitution $p_\mu \to 2\sin(p_\mu/2)$. For simplicity of notation
we shall continue to write $p_\mu$.

\begin{figure}[!htb]
   \begin{tabular}{cc}
      \includegraphics[width=6.50cm]{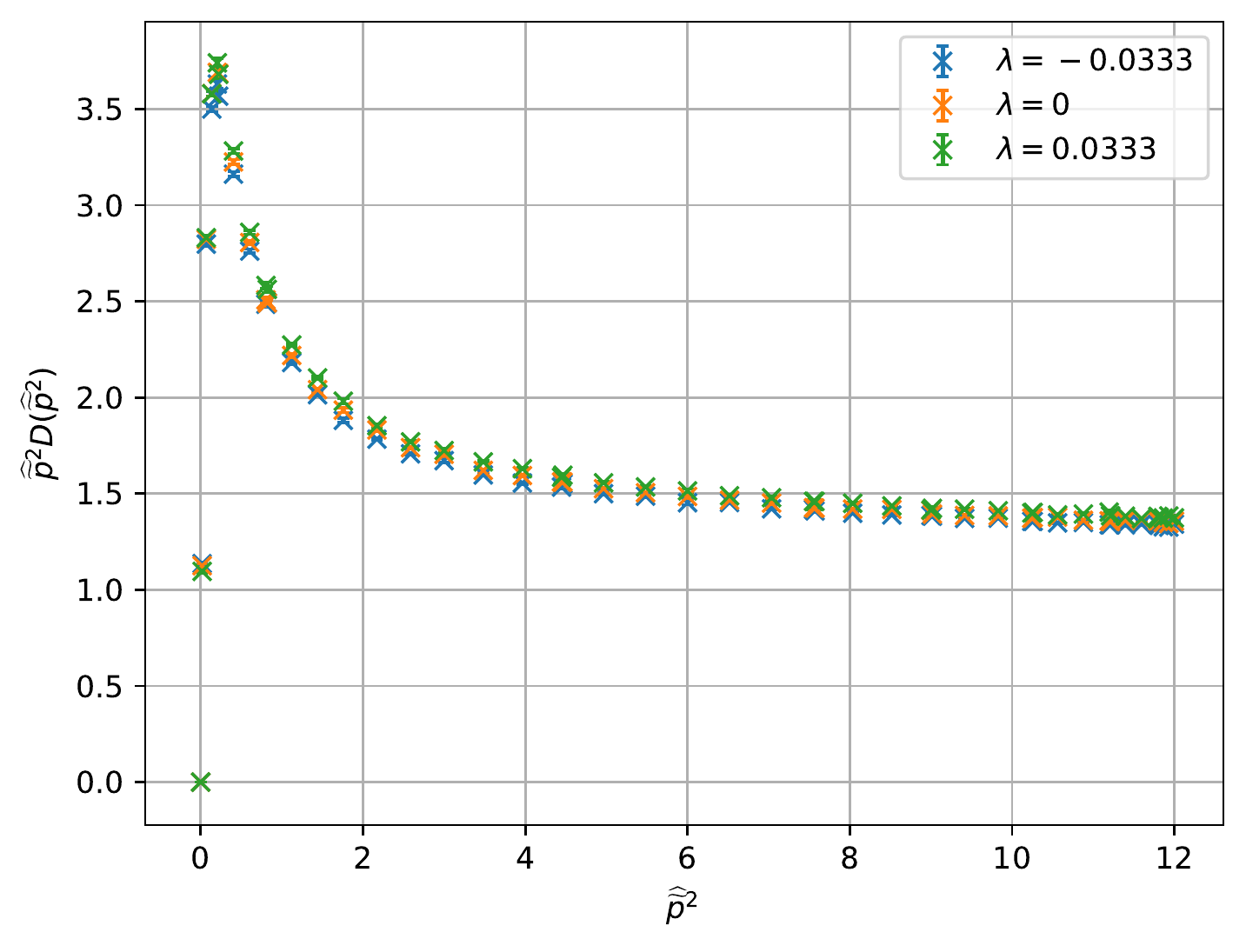}
      \hspace{0.50cm}
      \includegraphics[width=8.00cm]{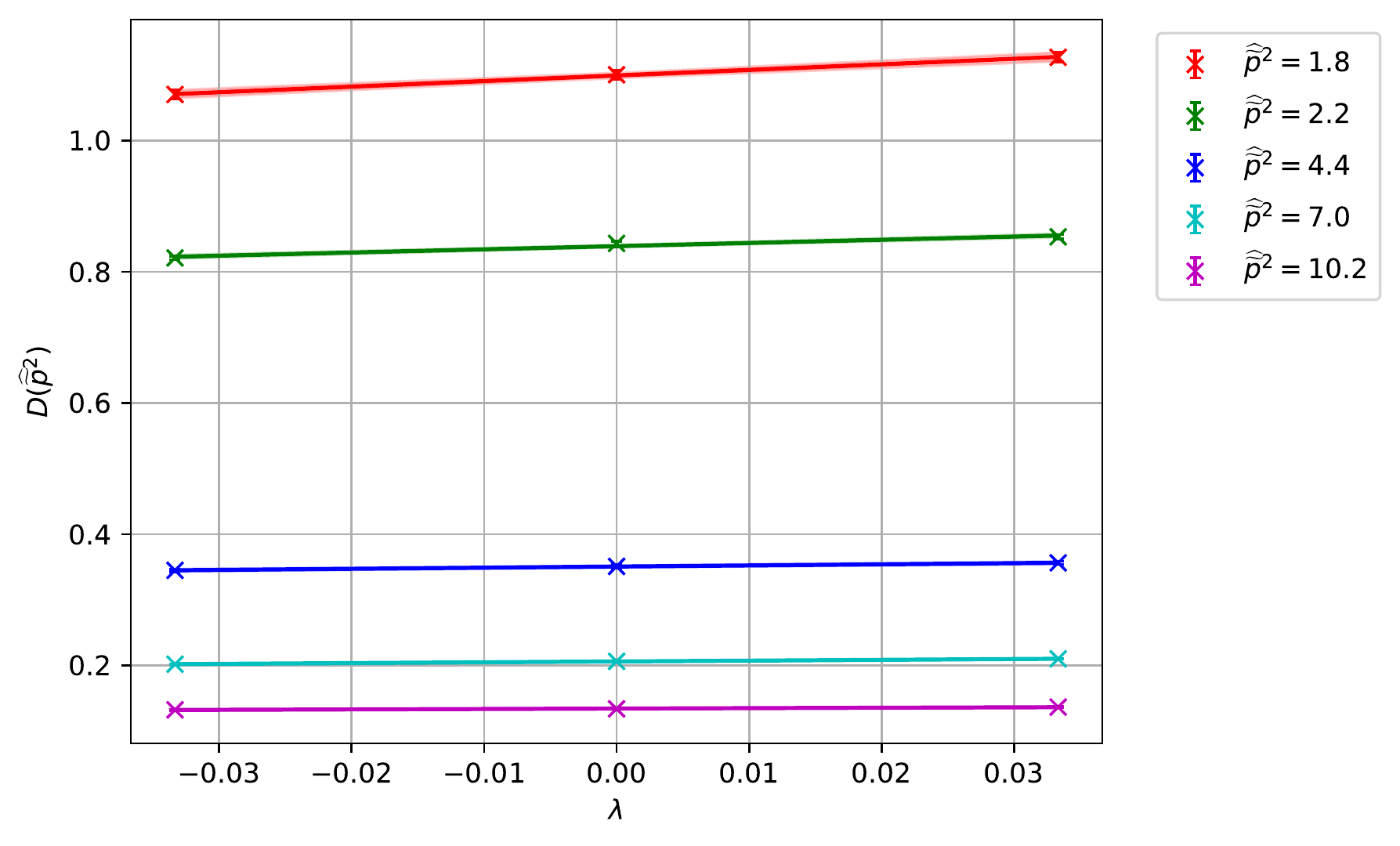}
   \end{tabular}
\caption{Left panel: $p^2\mbox{tr}\, D_\lambda(p)$ versus 
         $p^2$ for $\lambda = 0$, $\pm 0.0333$.
         Right panel: $\mbox{tr}\, D_\lambda(p)$ versus $\lambda$
         for selected values of $p^2$, as given in the figure
         together with a linear fit (in $\lambda$). 
         $O(1000)$ configurations per $\lambda$ value were generated.}
\label{D_lam_plot}
\end{figure}

\begin{figure}[!htb]
   \begin{center}
      \includegraphics[width=6.50cm]{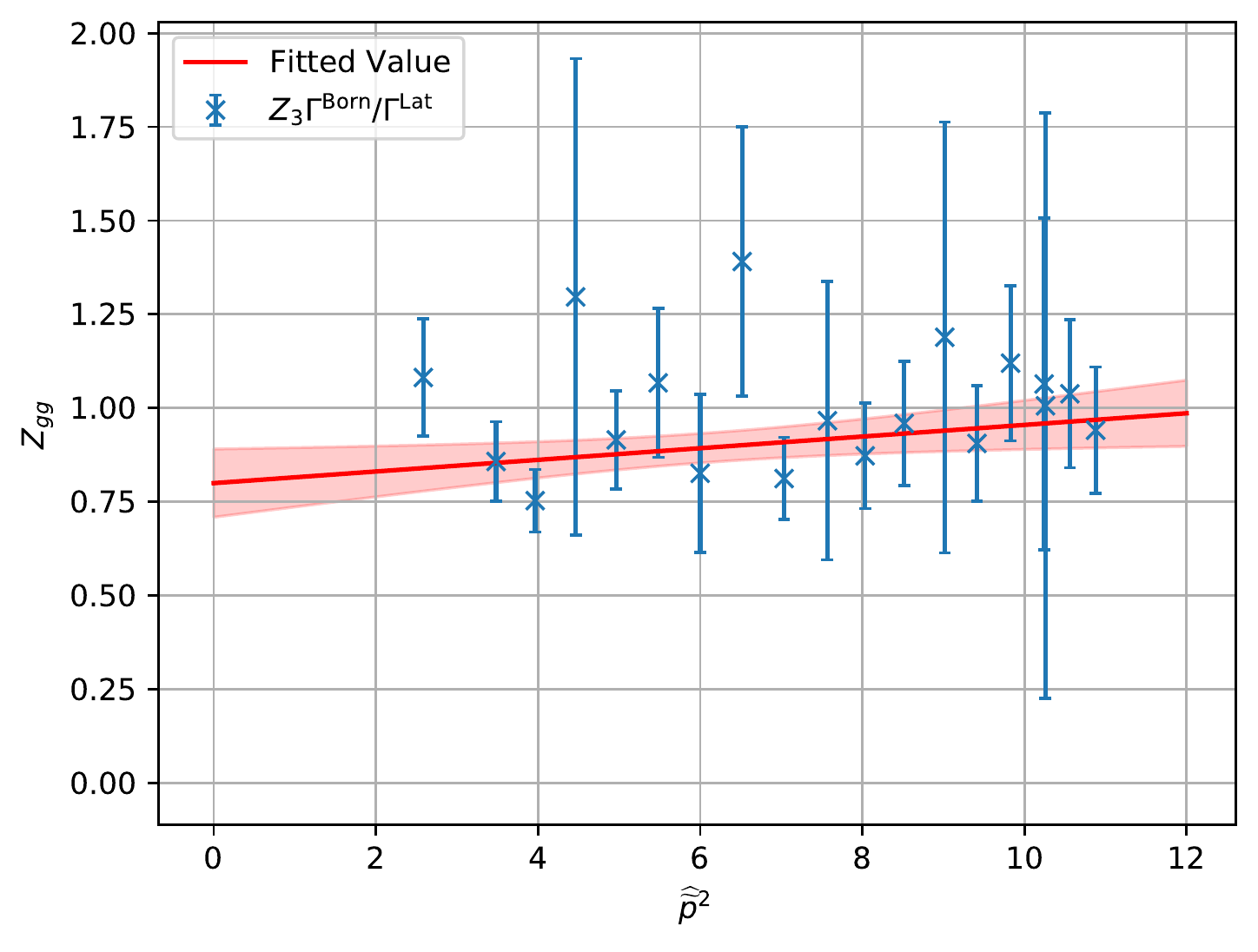}
   \end{center}
\caption{$Z_g$ as given in eq.~(\protect\ref{Z_g_I}) 
         versus $p^2$, together with a linear fit.}
\label{Z_g_plot}
\end{figure}

In Fig.~\ref{D_lam_plot} we plot $p^2\mbox{tr} \, D_\lambda(p)$ 
(for $\beta =6.0$ on a $24^3\times 48$ lattice in the Landau gauge,
$\xi = 0$) against $p^2$ where $p = (2\pi/24)(n,n,n,0)$ i.e.\ with 
a `cylinder' cut (left panel) and against $\lambda$ (right panel). 
From the gradients of the fits for each $p^2$ (some selected values
are given in the right panel of Fig.~\ref{D_lam_plot}) we can determine
$Z_g$ as given in eq.~(\ref{Z_g_I}). In Fig.~\ref{Z_g_plot} we 
we compare $Z_g$ determined from eq.~(\ref{Z_g_I}) together with
a linear fit $Z_g = A + Bp^2$, where the gradient term is taken 
to represent residual lattice effects.
We find $A = 0.799(91)$ for $Z_g$. Work is in progress to try to 
reduce the errors.

As a benchmark comparison from \cite{Engels:1999tk} we have
$Z_g |_{g^2=1} = (1 - 1.0225g^2 + 0.1305g^4)/(1 - 0.8557g^2)
                |_{g^2 = 1} = 0.748(20)$.
(This follows from setting $Z_g = 1 -g^2/2(c_\sigma-c_\tau)$, together
with a non-perturbative determination of the anisotropic coefficients 
$c_\sigma$ and $c_\tau$, see also \cite{Meyer:2007ic}.) This comparison
to \cite{Engels:1999tk} is the main result given here.

%----------------------------------------------------------------------------

\section{Conclusions and outlook}
\label{conclusions}

%----------------------------------------------------------------------------

In conclusion $\langle x \rangle_g^{\ind R}$ is a notoriously difficult quantity 
to compute as it is a short distance quantity with numerically large 
fluctuations -- it is a `disconnected quantity'. A straightforward 
determination requires hundreds of thousands of configurations. 
We have developed a FH technique, now including the renormalisation,
which although several runs are required each run is only moderately expensive.

We note that it is also possible to determine $Z_{gq}$ in the same way 
using the FH theorem after suitably modifying the quark propagator, when
\begin{eqnarray}
   Z_{gq} \propto \mbox{tr}\, \Gamma^{(b)\,{\ind Born}}
                           (S_{\lambda}^{-1} - S_0^{-1}) \,.
\end{eqnarray}
Finally a few comments about a more realistic computation with $2+1$ 
dynamical flavours of quarks. This has a more complicated renormalisation
pattern. We have the general structure
\begin{eqnarray}
   \left( \begin{array}{l}
             \langle x \rangle_g           \\[0.025in]
             \langle x \rangle_u^{\ind con}  \\
             \langle x \rangle_d^{\ind con}  \\
             \langle x \rangle_s^{\ind con}  \\[0.025in]
             \langle x \rangle_u^{\ind dis}  \\
             \langle x \rangle_d^{\ind dis}  \\
             \langle x \rangle_s^{\ind dis}  \\
             \langle x \rangle^{\ind con}_{q_v}  \\
          \end{array}
    \right)^{\ind R}
    = \left( \begin{array}{cccccccc} 
                Z_{gg} & Z_{gq} & Z_{gq} & Z_{gq} & Z_{gq} & Z_{gq} & Z_{gq} & Z_{gq}  \\[0.025in]
                  0   & Z_a-Z_b&  0    &   0   &   0   &   0  &   0    & Z_a-Z_b \\ 
                  0   &   0   & Z_a-Z_b&   0   &   0   &   0  &   0    & Z_a-Z_b \\ 
                  0   &   0   &  0     &Z_a-Z_b&   0   &   0  &   0    & Z_a-Z_b \\[0.025in]
                Z_{qg} & Z_b   & Z_b    & Z_b   & Z_a    & Z_b  & Z_b    & Z_b     \\
                Z_{qg} & Z_b   & Z_b    & Z_b   & Z_b    & Z_a  & Z_b    & Z_b     \\
                Z_{qg} & Z_b   & Z_b    & Z_b   & Z_b    & Z_b  & Z_a    & Z_b     \\[0.025in]
                  0   &   0   &  0     &   0   &   0   &   0  &   0    & Z_a-Z_b \\
              \end{array}
      \right) \,
      \left( \begin{array}{l}
                \langle x \rangle_g            \\[0.025in]
                \langle x \rangle_u^{\ind con}   \\
                \langle x \rangle_d^{\ind con}   \\
                \langle x \rangle_s^{\ind con}   \\[0.025in]
                \langle x \rangle_u^{\ind dis}   \\
                \langle x \rangle_d^{\ind dis}   \\
                \langle x \rangle_s^{\ind dis}   \\
                \langle x \rangle^{\ind con}_{q_v}\\
          \end{array}
    \right)^{\ind lat} \,,
\end{eqnarray}
where we consider here the case of $n_f = 3$ quarks and $n_{f_v} = 1$ 
partially quenched (or valence) quarks. Practically it is easier 
to split the terms into quark-line-connected and -disconnected pieces,
$\langle x \rangle_q = \langle x \rangle_q^{\ind con}
                               + \langle x \rangle_q^{\ind dis}$
with for example 
$\langle x \rangle_q^{{\ind con}\,{\ind R}}
     = Z_{qq}^{\ind NS}\langle x \rangle_q^{{\ind con}\,{\ind lat}}$.
All $Z$s depend on scheme and renormalisation scale $\mu$. 
The non-singlet (e.g.\ $\langle x \rangle_u - \langle x \rangle_d$)
and singlet (i.e.\ 
$\langle x \rangle_u+\langle x \rangle_d+\langle x \rangle_s$) 
renormalisation constants are thus
\begin{eqnarray}
   Z_{qq}^{\ind{NS}} = Z_a - Z_b\,, \qquad 
   Z_{qq}^{\ind{S}} = Z_{qq}^{\ind{NS}} + n_fZ_b \,,
\end{eqnarray}
respectively.
%Bottom row of zeros:
%         if don't put in a valence $\langle x \rangle_v$ `by hand'
%         then it remains zero
%Last column:
%         valence quark surrounded by cloud of $g$,
%         $u - \overline{u}, \ldots$ quark bubbles
%Valence means only for `quark line connected',
%         so renormalise as
%         $\langle x \rangle_v^{\ind{R}}
%           = Z_{qq}^{\ind{NS}}\langle x \rangle_v^{\ind{LAT}}$ 
As before we have
\begin{eqnarray}
   \left(\langle x \rangle_g +
         \sum_q \langle x \rangle_q + \sum_{q_v} \langle x \rangle_{q_v}
   \right)^{\ind R}
      &=& Z_g \langle x \rangle_g^{\ind lat}
          + Z_q\left( \sum_q \langle x \rangle_q
                + \sum_{q_v} \langle x \rangle_{q_v}\right)^{\ind lat}
       = 1 \,,
                                                             \nonumber
\end{eqnarray}
giving
\begin{eqnarray}
   Z_g = Z_{gg}^{\ind \overline{MS}} + n_fZ_{qg}^{\ind \overline{MS}}\,,
   \qquad
   Z_q = Z_{gq}^{\ind \overline{MS}} + Z_{qq}^{{\ind NS} \,\ind \overline{MS}} \,,
                                                             \nonumber
\end{eqnarray}
with as before $Z_g$, $Z_q$ just depending on the coupling, $g$,
but individual terms depend on the chosen scheme, e.g.\ $\overline{MS}$.
A similar FH scheme for the renormalisation is being developed here.

%----------------------------------------------------------------------------

\section*{Acknowledgements}

%----------------------------------------------------------------------------

The numerical configuration generation (using the BQCD lattice 
QCD program \cite{Haar:2017ubh})) and data analysis 
(using the Chroma software library \cite{edwards04a}) was carried out
on the IBM BlueGene/Q and HP Tesseract using DIRAC 2 resources 
(EPCC, Edinburgh, UK), the IBM BlueGene/Q (NIC, J\"ulich, Germany) 
and the Cray XC40 at HLRN (The North-German Supercomputer 
Alliance), the NCI National Facility in Canberra, Australia 
(supported by the Australian Commonwealth Government) 
and Phoenix (University of Adelaide).
RH was supported by STFC through grant ST/P000630/1.
HP was supported by DFG Grant No. PE 2792/2-1.
PELR was supported in part by the STFC under contract ST/G00062X/1.
GS was supported by DFG Grant No. SCHI 179/8-1.
RDY and JMZ were supported by the Australian Research Council Grant
No. DP190100297. We thank all funding agencies.

% ------------------------------------------------------------------

%----------------------------------------------------------------------------

\end{document}